\documentstyle[12pt]{article}
\title{Non-periodic long-range order for fast decaying interactions
at positive temperatures}
\author{Aernout C. D. van Enter \\ 
Institute for Theoretical Physics \\
Rijksuniversiteit Groningen   \\
Nijenborgh 4, 9747 AG Groningen \\
THE NETHERLANDS   \\
aenter@phys.rug.nl \\ [5mm]
Jacek Mi\c{e}kisz   \\
Institute of Applied Mathematics and Mechanics \\
Warsaw University \\
ul. Banacha 2, 02-097 Warsaw \\  
POLAND \\
miekisz@mimuw.edu.pl  \\ [5mm]
Milo\v{s} Zahradn\'\i k \\
Faculty of Mathematics and Physics \\
Charles University Prague \\
Sokolovsk\'a 83, 18600 Prague \\
CZECH REPUBLIC           \\
mzahrad@karlin.mff.cuni.cz}    
\begin{document}
\maketitle
\baselineskip=14pt
\noindent {\bf Abstract.} We present the first example
of an exponentially decaying interaction which gives rise to non-periodic
long-range order at positive temperatures.
\eject
\newtheorem{theorem}{Theorem}          
\newtheorem{lemma}[theorem]{Lemma}              
\newtheorem{proposition}[theorem]{Proposition}
\newtheorem{corollary}[theorem]{Corollary}
\newtheorem{definition}[theorem]{Definition}
\newtheorem{conjecture}[theorem]{Conjecture}
\newtheorem{claim}[theorem]{Claim}
\newtheorem{observation}[theorem]{Observation}
\def\proof{\par\noindent{\it Proof.\ }}
\def\reff#1{(\ref{#1})}

\let\zed=\bbbz 
\let\szed=\bbbz 
\let\IR=\bbbr 
\let\R=\bbbr 
\let\sIR=\bbbr 
\let\IN=\bbbn 
\let\IC=\bbbc 

\def\nl{\medskip\par\noindent}

\def\scrb{{\cal B}}
\def\scrg{{\cal G}}
\def\scrf{{\cal F}}
\def\scrl{{\cal L}}
\def\scrr{{\cal R}}
\def\scrt{{\cal T}}
\def\pfin{{\cal S}}
\def\prob{M_{+1}}
\def\cql{C_{\rm ql}}
\def\bydef{\stackrel{\rm def}{=}}   
\def\qed{\hbox{\hskip 1cm\vrule width6pt height7pt depth1pt \hskip1pt}\bigskip}
\def\remark{\medskip\par\noindent{\bf Remark:}}
\def\remarks{\medskip\par\noindent{\bf Remarks:}}
\def\example{\medskip\par\noindent{\bf Example:}}
\def\examples{\medskip\par\noindent{\bf Examples:}}
\def\nonexamples{\medskip\par\noindent{\bf Non-examples:}}

\newenvironment{scarray}{
          \textfont0=\scriptfont0
          \scriptfont0=\scriptscriptfont0
          \textfont1=\scriptfont1
          \scriptfont1=\scriptscriptfont1
          \textfont2=\scriptfont2
          \scriptfont2=\scriptscriptfont2
          \textfont3=\scriptfont3
          \scriptfont3=\scriptscriptfont3
        \renewcommand{\arraystretch}{0.7}
        \begin{array}{c}}{\end{array}}

\def\wspec{w'_{\rm special}}
\def\mup{\widehat\mu^+}
\def\mupm{\widehat\mu^{+|-_\Lambda}}
\def\pip{\widehat\pi^+}
\def\pipm{\widehat\pi^{+|-_\Lambda}}
\def\ind{{\rm I}}
\def\const{{\rm const}}

\bibliographystyle{plain}

\section{Introduction}
Since the discovery of quasicrystals \cite{SBGC}, there has been an interest
in understanding their occurrence in statistical mechanics models of
interacting particles, see for example \cite{Mi1,BQ,Rad1,Sen}. One would 
like to show that a quasicrystalline phase occurs in appropriate
models at sufficiently low temperatures. We interpret this as the
occurrence of ground states or Gibbs states which possess a quasi-periodic, 
or more generally, a non-periodic long-range order \cite{EM2,Rue1,Aub1,rad}.

Up till now there only exist some partial results going
in this direction. Most of them have been obtained for lattice models,
and here again we will get a result of this type.

In classical lattice-gas models, every site of a regular lattice, $Z^{d}$,
is occupied by one of $n$ different types of particles (equivalently,
by $+1$ or $-1$ in spin 1/2 models). Configurations of such models are
therefore elements of $\Omega=\{1,...,n\}^{Z^{d}}$. Particles interact
through possibly many-body potentials which are represented by
functions $\Phi_{\Lambda}:\Omega_{\Lambda} \rightarrow R$ for all finite
$\Lambda \subset Z^{d}$, where $\Omega_{\Lambda}=\{1,...,n\}^{\Lambda}$. 
We assume that the $\Phi_{\Lambda}$ are
translation invariant and decay exponentially in $N(\Lambda)$, the number of sites in $\Lambda$, and in fact in  what we need for the next section, even in $diam (\Lambda)$.
The formal Hamiltonian can be therefore written as
$H_{\Phi}=\sum_{\Lambda}\Phi_{\Lambda}.$ By ground states of $H_{\Phi}$
we mean translation-invariant probability measures supported by 
configurations with minimal energy density. Ground states are
zero-temperature limits of translation-invariant Gibbs states
(equilibrium states).
\bigskip

\noindent Following are the main results which are known to us, 
concerning non-periodic order
of ground states and Gibbs states of lattice models:  
\smallskip

\noindent 1) For finite-range interactions, non-periodic long-range order
in the ground state can occur in dimension 2 and higher,
but not in dimension 1 \cite{Rad1,RadSchu,mier2}.  
\smallskip

\noindent 2) In dimension 1, non-periodic long-range order
in the ground state can occur for infinite-range, but arbitrarily
fast decaying interactions \cite{GERM,Aub1,rad}.
\smallskip
 
\noindent 3) In dimension 2, non-periodic long-range order
in the ground state can
occur for nearest-neighbor interactions \cite{rad2,mier,Rad1,Mi1,mie3,mie4}.
At positive temperatures in such models, the best result proven
so far is the existence of an infinite sequence of temperatures
with the period doubling of periodic Gibbs states \cite{mie2}. 
\smallskip

\noindent 4) At positive temperatures, non-periodic long-range order can
occur for slowly decaying (summable) interactions
(in arbitrary dimensions). These interactions can be finite-body, or 
even pair interactions \cite{EM1,EZ}.
\bigskip

\noindent In this note we want to present an example
and a general construction where non-periodic
long-range order occurs at positive temperatures for fast 
(exponentially) decaying interactions in 3 dimensions.
Before entering into details of the argument, which is based on 
\cite {HoZa}, and is indeed more or less a corollary of that paper, 
we want to make a few comments. 
\vspace{2mm}

\noindent 1) Although at zero temperature
there exists an important qualitative
difference between strictly finite-range interactions and interactions
which are of infinite range but have a fast decaying tail, this qualitative
difference is not expected to persist at positive temperatures. 
\smallskip

\noindent 2) There is a conjecture that two-dimensional lattice gas models
with short-range interactions always have at most
finitely many (periodic) Gibbs states, 
which rules out the possibility  of non-periodic
long-range order. For some arguments and results supporting this 
conjecture, see \cite{Sin,Sla,DobShl1}. It would mean that our results
could not be true in 2 dimensions. 
\smallskip
 
\noindent 3) A limitation of our result is that we prove
the existence of non-periodic
structures in only one direction. We conjecture that this limitation
is not really necessary, but our method does not give the stronger
result (non-periodic order in three directions at the same time).

\section{The example}
Let us recall a definition of the Thue-Morse state.
We begin by constructing a one-sided Thue-Morse sequence.
We put $+$ at the origin and perform successively a substitution:
$+ \rightarrow +-, - \rightarrow -+$, obtaining $+, +-, +--+, +--+-++-, ...$
We get a one-sided sequence $\{X_{TM}(i)\}, i \geq 0$. We define
$X_{TM} \in \Omega=\{+,-\}^{Z}$ by setting $X_{TM}(i)=X_{TM}(-i-1)$ for $i<0.$

Let $T$ be a translation operator, i.e., $T:\Omega \rightarrow \Omega$,
$T(X)(i)=X(i-1)$, $x \in \Omega$. Let $G_{TM}$ be the closure (in the product
topology of the discrete topology on $\{+,-\}$) of the orbit
of $X_{TM}$ by translations, i.e., $G_{TM}=\{T^{n}(X_{TM}), n \geq 0\}^{cl}$.
It can be shown that $G_{TM}$ supports exactly one translation-invariant
probability measure $\mu_{TM}$ on $\Omega$ \cite{kak,kea}. $\mu_{TM}$ 
is the only ground state of a certain fast decaying four-spin interaction
\cite{GERM} (see (1) below).

We will combine this construction of \cite{GERM} with the result of
\cite{HoZa} on ``stratified'' Gibbs measures. Thus in one of the directions
we start by choosing the interaction of \cite{GERM},
while in the other two directions we have the nearest-neighbor ferromagnetic
interaction. This model has as ground-state configurations the ``stratified''
(or stacked) Thue-Morse sequences, that is they are 
Thue-Morse sequences (i.e. elements of $G_{TM}$) in one direction,
and translation-invariant in the other two directions.
Thus in the terminology of \cite{HoZa} we are in a stratified situation:
even though the interaction is translation invariant, ground-state
configurations building up the single translation-invariant
ground-state measure are nonperiodic (Thue-Morse) layered structures.

Let us mention some other works considering low-temperature behavior
of models with periodic stratified ground states. 
The phase diagram of the ANNNI
model was investigated in \cite{sel,sin1,sin2}. General three-dimensional
stratified models were discussed in \cite{fish} and some related
results were recently obtained in a model with a layered magnetic
field \cite{laan}. 
 
It has been known for some time that ground-state configurations which have no
energy barrier between them in one dimension (example: the kink states
in the one-dimensional Ising model) can give rise to corresponding 
Gibbs states in three dimensions. These Gibbs states have the same
structure in one direction and are ferromagnetically ordered in the 
other two directions (example: the Dobrushin states in the three-dimensional
Ising model \cite{Dob1}). The ideas of Dobrushin have been extended to
more general situations in \cite{HoZa}, and here we observe that these
recent results can be applied to the non-periodic Thue-Morse ground-state
configurations studied in
\cite{GERM}. In one dimension, we will consider the
sequences supporting the translation-invariant Thue-Morse measure $\mu_{TM}$ 
(which is the only translation-invariant ground-state measure).
The energy barriers between them
may be arbitrarily small, but this does not need to matter; in fact, as we
have just remarked, even for ground states with zero-energy barriers, 
it is the case that adding ferromagnetic nearest-neighbor terms 
in two extra dimensions can stabilize them. In the general approach
of \cite{HoZa}, the result is that the stratified structures appear,
not necessarily for the original interaction, but for a small (weak    
and exponentially decaying) perturbation thereof. In our case
it is to be expected that in fact such a perturbation will be needed.
We will start with the Hamiltonian $H_{TM}+H_{F}$, where
\begin{equation}
\ H_{TM}=\sum_{i \in Z^{3}} \sum_{r=0}^{\infty} \sum_{p=0}^{\infty}
J(r,p)(\sigma_{i} + \sigma_{i+(2^{p})e_{1}})^{2} (\sigma_{i+(2r+1)2^{p}e_{1}}
+ \sigma_{i+(2r+2)2^{p}e_{1}})^{2},
\end{equation}
\begin{equation}
H_{F}= \sum_{i \in Z^{3}} J(\sigma_{i} \sigma_{i+e_{2}} +
\sigma_{i} \sigma_{i+e_{3}}),
\end{equation}
where $e_{n}$ is the unit vector in the $n-th$ direction, $\sigma_{i}=\pm 1$
is a spin variable and $J(r,p)>0$ and decays to zero exponentially fast
when distances between interacting particles increase. 

The Thue-Morse layers are ground-state configurations of $H_{TM}+H_{F}$.
We need to take care against the possibility that there are periodically
layered structures which have more low-energy excitations and therefore
compensate for having a higher energy at zero temperature. To illustrate
the phenomenon we need to control, assume for the moment
that in the first sum, only the term $J(0,0)$ in the above expression is 
non-zero. That is, one excludes local configurations with three successive 
pluses or minuses. Then, in the horizontal direction,
next to the Thue-Morse sequences, there are many more ground states, and it is
easy to see that, for example, the 3-periodic structures $--+$ or $++-$ have
more lowest energy excitations ($8J$ which is the energy of overturning 
one spin without creating three successive pluses or minuses) 
than (and hence in the terminology of \cite{brisla} dominate) 
the Thue-Morse ones. This effect explains the necessity of having 
an extra interaction term in our theorem, to make low-temperature 
Gibbs measures ``Thue-Morse-like'' in the first 
direction.

In the following, we will consider one-dimensional interactions
$\Phi=\{\Phi_{\Lambda}: \Omega_{\Lambda} \rightarrow R\}$ such that 
for every $X \in \Omega_{\Lambda}$,
$|\Phi_{\Lambda}(X)| \leq \epsilon \omega^{diam(\Lambda)}$
for some $\epsilon, \omega >0$. We denote by ${\cal H}^{\epsilon, \omega}$
the family of such interactions. If $X$ is a ground-state configuration,
then by a Gibbs state which is a small perturbation of it we mean a Gibbs state
$\rho_{X}$ such that $\rho_{X}(P_{a}^{X})<\epsilon(T),$ where $P_{a}^{X}$
is a projection on configurations which are different from $X$
at a lattice site $a$ and $\epsilon(T) \rightarrow 0$, when the temperature
$T \rightarrow 0$.

The subsequent theorems follow from the Main Theorem of \cite{HoZa}. Proofs and
more technical details one can find there.
\begin{theorem}
Let $H=H_{0}+H_{F}$, where $H_{0} \in {\cal H}^{\epsilon_{0}, \omega}$,
$\epsilon_{0} << 1$, $\omega=Ce^{-2J/T}$ for some constant $C$, and
$J$ appears in (2). Then there is $H_{1}
\in {\cal H}^{\epsilon_{1}, \omega}$, where $\epsilon_{0} < \epsilon_{1}<
\epsilon_{0}+\omega$
such that if $X$ is a ground-state configuration of $H^{*}=H_{1}+H_{F}$,
then there exists a Gibbs state of $H$, $\mu_{X}$, which is a small
perturbation of $X$.
\end{theorem} 
Actually, one can give an explicit formula for the perturbative Hamiltonian
$H_{1}-H_{0}$ in terms of quickly converging cluster expansion series, whose
(small!) terms change only slowly with $H_{0}$ and $H_{F}$.

Our aim is to construct a translation-invariant Gibbs state of $H$
such that it has only non-periodic Gibbs states in its extremal
decomposition. In order to do so we would like $H^{*}$ to have
a unique ground-state measure supported by non-periodic ground-state
configurations (Thue-Morse stratified sequences in our example). The following inverse
mapping type theorem assures us of this (the Lipschitz property is rather
obvious if one inspects the explicit formulas given in \cite{HoZa} 
for $H_{1}$).
\begin{theorem}
The above mapping $H \rightarrow H^{*}$ is Lipschitz continuous
in the sense that if $H_{0}-H_{0}' \in {\cal H}^{\epsilon', \omega}$
with $\epsilon'< \epsilon$, then $H_{1}-H_{1}' \in
{\cal H}^{\epsilon'', \omega}$ with $\epsilon'' = \epsilon'\omega$.
In particular, if we choose $\epsilon_{1}$ such that
$\epsilon_{0}-\epsilon_{1} > \omega$, then for every
$H^{*} \in {\cal H}^{\epsilon_{1}, \omega}$,
there exists a preimage $H \in {\cal H}^{\epsilon_{0}, \omega}$
and such that there is one-to-one correspondence between
stratified ground-state configurations of $H^{*}$ and stratified Gibbs
states of $H$.
\end{theorem}
Now we put $H^{*}=H_{TM}+H_{F}$ and use Theorem 1 and 2 to obtain
nonperiodic Thue-Morse Gibbs states of $H$, which are small perturbations
of the Thue-Morse stratified ground-state configurations. 
Therefore, there exists a translation-invariant Thue-Morse Gibbs
state, $\rho_{TM}$, which has only nonperiodic (Thue-Morse)
Gibbs states in its extremal decomposition. 

$\rho_{TM}$ is a Gibbs state which is extremal among
the translation-invariant Gibbs states of $H$. We expect 
that $H$ does not have any other translation-invariant Gibbs states.
However, we cannot exclude at the moment some ``exotic'' translation-invariant
states which do not arise from stratified configurations. 

\section{Generalizations and open problems}
It was shown by Aubry and Radin \cite{Aub1,rad} that any strictly
ergodic measure on $\Omega$ is a unique ground state of a certain
one-dimensional, many-body, infinite-range but  arbitrarily 
fast decaying interaction. Our construction shows the existence of another 
interaction of the same type such that when adding ferromagnetic 
interactions in two extra dimensions one obtains a model with 
a low-temperature Gibbs state which is a small perturbation 
of the original measure.

As we have already discussed, we do not expect the nonperiodic order
for the original Thue-Morse interaction. A likely possibility
here is the existence of an infinite sequence of
temperatures, decreasing to zero,  at which the periods of the corresponding  
extremal 
Gibbs states grow. It is an 
open problem to construct a two-body (or even finite-body) interaction 
(finite range or
exponentially decaying) without periodic ground-state configurations
and with a nonperiodic Gibbs state. 
\bigskip

\noindent {\em Acknowledgements}:
We thank Peter Holick\'y for his contributions to the stratified theory, and for some useful conversations.
A.C.D. v. E. and J. M. thank the Charles University of Prague for hospitality.
This research has been supported by EU contract CHRX-CT93-0411 and
CIPA-CT92-4016.
J.M. would like to thank the Polish Scientific Committee for Research, 
for a financial support under the grant KBN 2 P03A 015 11. M.Z. was
partially supported by Czech Republic grants 202/96/0731 and
96/272.

\end{document}